\documentclass[sigconf,screen]{acmart}
\AtBeginDocument{%
  }

\copyrightyear{2026}
\acmYear{2026}
\setcopyright{cc}
\setcctype{by}
\acmDOI{XXXXXXX.XXXXXXX}
\acmConference[CHI '26]{the 2026 CHI Workshop on Data Literacy for the 21st Century: Perspectives from Visualization, Cognitive Science, Artificial Intelligence, and Education}{April 13, 2026}{Barcelona, Spain}
\acmISBN{978-1-4503-XXXX-X/2026/04}

\usepackage{microtype}

\begin{document}

\title{Towards Measuring Interactive Visualization Abilities: Connecting With Existing Literacies and Assessments}


\author{Gabriela Molina Le\'{o}n}
\orcid{0000-0002-9223-2022}
\affiliation{%
  \institution{Aarhus University}
  \city{Aarhus}
  \country{Denmark}}
\email{leon@cs.au.dk}

\author{Benjamin Bach}
\orcid{0000-0002-9201-7744}
\affiliation{%
  \institution{Inria Bordeaux}
  \city{Bordeaux}
  \country{France}}
\email{benjamin.bach@inria.fr}

\author{Matheus Valentim}
\orcid{0000-0003-0860-2084}
\affiliation{%
  \institution{Aarhus University}
  \city{Aarhus}
  \country{Denmark}}
\affiliation{%
  \institution{University of Illinois}
  \city{Urbana-Champaign}
  \state{IL}
  \country{USA}}
\email{matheus7@illinois.edu}

\author{Niklas Elmqvist}
\orcid{0000-0001-5805-5301}
\affiliation{%
  \institution{Aarhus University}
  \city{Aarhus}
  \country{Denmark}}
\email{elm@cs.au.dk}

\renewcommand{\shortauthors}{Molina Le\'{o}n et al.}

\begin{abstract}
How do we assess people's abilities to interact with data visualizations?
The current state-of-the-art visualization literacy tests---such as VLAT and its derivatives---only involve the use of static visualizations.
Despite advances in investigating multiple visualization abilities, we do not yet have formal methods to assess the ability of a person to interact with a data visualization effectively.
In this position paper, we discuss related literacy concepts and assessments to propose and compare different approaches for assessing the abilities that people leverage to use visualizations in interactive sensemaking tasks.
\end{abstract}



\maketitle

\section{Introduction}

A person's ability to read and create visualizations is referred to as their \textit{visualization literacy}~\cite{boy14}, but the prevalent understanding of this concept is still an evolving and active area of research in data visualization.
To characterize and understand the abilities involved in \textit{interacting} with visualizations, we recently proposed a theoretical model to define and describe \textit{interactive visualization literacy} (IVL)~\cite{leon26IVL}. 
The model describes the visualization action cycle through three levels of abstraction and three interaction gulfs, identifying nine different \textit{literacies} that people leverage to interact with visualization systems, including the standard visualization literacy. 
However, there are still no visualization assessments that measure interaction-related abilities. 
State-of-the-art assessment tests, such as VLAT~\cite{lee17} and its derivatives, only measure a person's ability to use a static visualization.
Thus, the next step is to investigate how the interactive abilities captured in the IVL model can be assessed, which requires designing and developing ways to effectively measure them.

Interacting with data visualizations goes beyond interpreting visual elements.
It requires setting goals and executing actions, potentially in an iterative manner, until reaching the desired visualization state.
This complexity is nicely illustrated by the statement of one participant in the observational study conducted to scrutinize the IVL model~\cite{leon26IVL}:

\begin{quote}
    \textit{``Even though we explained that I would be interacting with this data, I… kind of forgot about it, and I think I'm used to just, you know, looking at it.
    Not having to do something for it to give me information, it's just, you know, there.
    But after I understood that I could click and new information would be given, it was nice, but… still, it was kind of hard to understand what information would be given depending on how I interacted.''} --- P8
\end{quote}

The statement reflects how using an interactive visualization is perceived differently from using a static visualization.
To measure interaction abilities, we need to go beyond visual interpretation and consider interface design, interaction techniques, input devices, and potentially other aspects that we have yet to identify.

In this position paper, we argue that, in order to assess people's interaction abilities, we should draw from existing literacies and assessments that have been applied or discussed in related domains.
We discuss different literacy concepts and assessments from the human-computer interaction (HCI) and the education domains, which we use as a base to identify and compare multiple assessment approaches inspired by said work.

\section{Interaction with Data Visualizations}
\label{sec:interaction}

If visualization is the interactive graphical representation of data to amplify cognition~\cite{card99}, then its practical application is the interactive use of such graphical representations for data exploration~\cite{DBLP:books/lib/Tukey77}, sensemaking~\cite{pirolli2005sensemaking}, and decision making~\cite{DBLP:journals/tvcg/DimaraZTF22}.
In interactive visualization systems, we can identify two major flows of information. 

One flow is from data to a visual representation via the intermediary steps of analysis and visual mapping.
On a technical level, this sequential transformation of data from symbolic into geometric form is what has been called the \textit{visualization pipeline}~\cite{card99} (or the \textit{visualization reference model}~\cite{DBLP:conf/infovis/Chi00}). 
However, as has been noted by many authors, the visualization pipeline is really an interactive system where each stage of the pipeline supports user input for, e.g., data wrangling~\cite{DBLP:journals/ivs/KandelHPKHRWLBB11}, filtering and transformation~\cite{yi07}, and view manipulation and navigation~\cite{yi07, DBLP:conf/infovis/AmarES05}. 

The other flow of information is from a person via some interactions to the interface and back again.
Any user interaction can alter parameters of the visualization pipeline if the system allows.
Interaction between a person and a visualization is enabled through a structure commonly called the \textit{user interface}, which at the same time defines the boundary between the person and the visualization system.
However, in the eyes of the user, the notions of ``user interface'' and ``visualization'' are often blurred.
This double notion is most evident in the case of direct manipulation~\cite{DBLP:journals/computer/Shneiderman83}, semantic zoom~\cite{Bederson1994}, or any other interaction technique directly performed onto the visualization changing its appearance (visual mapping, rendering, styling, etc.): a visualization \textit{becomes} the interface (or \textit{instrument}~\cite{beaudouinlafon00}).
Still, the difference is necessary to separate purely graphical means of a visualization, i.e., visual marks, from its interaction affordances. 

\begin{figure*}[ht]
    \centering
    \includegraphics[width=\linewidth]{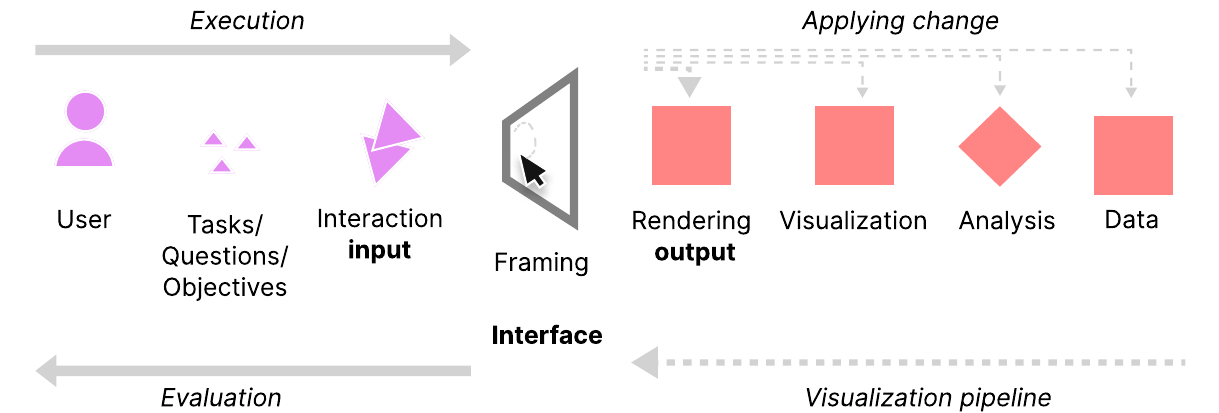}
    \caption{\textbf{Components of an interactive visualization system.} 
    From left to right, the visualization pipeline, the interface, and a user, their tasks, and interactions are embedded into the execution and evaluation.}
    \Description{An illustration showing the components. On the left, there are three elements: the user as a human figure, the tasks symbolized by smaller triangles, and the interaction input symbolized as big triangles. In the middle, a rectangle represents the framing. On the right, four rectangles represent the rendering output, the visualization, the analysis, and the data. Around all those elements, multiple arrows represent the direction of interaction flows from and towards the framing.}
    \label{fig:components}
\end{figure*}

We refer to the interface of a visualization system as composed of three parts, illustrated in \autoref{fig:components}: 

\begin{enumerate}
    \item \textit{output capabilities} (e.g., display) for rendering a visualization;
    \item \textit{input capabilities} to allow interacting with the system; and 
    \item a \textit{framing}, which are interface components (e.g., sliders), separate from the visualization itself.
\end{enumerate}

\section{Established Literacies and Assessments}
\label{sec:established}

While we believe that measurements of interaction abilities with visualization should be informed by prior work on visualization and interaction design, assessing abilities is not an endeavor exclusive to the visualization research field.
Beyond well-known assessments dedicated to static visualizations (e.g., VLAT~\cite{lee17}, CALVI~\cite{ge23}), many researchers and organizations have defined or applied other literacies and assessments that are related, depend on, or precede visualization abilities.
For instance, when researchers discuss \textit{data literacy}, they often associate it with visualizations~\cite{cui23}.
Shreiner~\cite{shreiner19} defined data literacy as \textit{``the ability to analyze, interpret, evaluate, and use data and data visualizations''} and examined it qualitatively, through a think-aloud exercise and semi-structured interviews with school students.
According to her, using visualizations is one of the multiple aspects that compose data literacy.
Conversely, there are other abilities, such as numeracy, that are closely related to (static) visualization literacy~\cite{lee19corr}. 
If we think of the foundations of visual perception, we should also consider the ability to perceive and interpret any visual artifacts (i.e., visual literacy), whose assessment has affective, compositional, and critical dimensions~\cite{callow08}.

The relationships among these concepts suggest that abilities are often interconnected, and assessment tools for one ability may serve as building blocks to measure another or complement other methods. 
Thus, advancements in interactive visualization assessments will benefit from identifying related abilities and their corresponding assessments.

\subsection{Incorporating Interaction Concepts}

Interacting with data visualizations requires input devices, modalities, updates, and reinterpretation.
Thus, assessing the abilities people use to interact will require considering these aspects that have not been addressed. 
Yet, there is work from other domains that has explored some of them and can serve as a starting point.
For example, in the context of Information and Communication Technology for Development (ICT4D), Wyche et al.~\cite{wyche16} proposed one way to examine device literacy, focusing on the everyday use of mobile phones. 
The researchers applied a mixed-methods approach to investigate the effectiveness of video material to improve the device literacy of farmers in Kenya.
They assessed device literacy through interviews and Likert scale questions (e.g., confidence level 1---3 about sending an SMS) to compare the answers of the farmers before and after being exposed to the explanatory videos. 
Similar approaches could help understand the abilities people develop with specific input and output devices.

In the psychology domain, Carolus et al.~\cite{carolus23} investigated the abilities people need for \textit{literate interactions} with voice-based AI systems.
From expert interviews, the authors derived 10 dimensions and integrated them into a digital interaction literacy model.
Several dimensions are relevant to any form of interaction, such as the need to understand how the input device works and people's ability to reflect on their interactive experiences to identify needs and adjust usage.
Taking these dimensions into account could help distinguish and isolate the dimensions of interaction abilities.

\subsection{Large-scale Assessments}

While most assessment tests in visualization research have been developed in collaboration with experts or through crowdsourcing experiments (e.g.,~\cite{cabouat25, ge23, pandey23}), there is a larger space for assessments at the level of international education policy.
International organizations, such as the Organisation for Economic Co-operation and Development (OECD), conduct assessment tests in dozens of countries, every few years, to make policy recommendations to state governments at an international level.
Learning and applying concepts from these established assessments can facilitate not only differentiating interaction abilities from well-known concepts but also developing robust tests.

For example, the Trends in International Mathematics and Science Study (TIMSS) is an assessment program of fourth- and eighth-grade students, conducted in at least 64 countries. 
The analysis differentiates between low, medium, and high digital self-efficacy, and in its 2023 cycle~\cite{vonDavier24}, the test assesses students' digital self-efficacy with response items related to visualization authoring, such as: \textit{``I can create tables, charts, and graphs using a computer, tablet, or smartphone.''} 

While international organizations usually focus on school students, assessments for adults also exist.
The OECD surveys adult skills~\cite{oecdadult25}, which include not only standard literacy and numeracy, but also problem-solving abilities, which are necessary to set goals and to develop strategies while interacting.
The test also considers visualization interpretation: adults with level-four skills can \textit{``read and interpret multi-variate data presented in a single graph.''} 

\autoref{tab:items} shows adapted items from existing assessments that we deem relevant to visualization interaction. 
Combining items from existing assessment tools with new tasks related to interactive visualizations could help us distinguish among the different characteristics or dimensions that underlie interaction abilities.

A potential issue with large-scale tests is that they are becoming fully digital, which may lead to additional challenges because people need to interact with a computer to answer.
Still, the interactive features of these assessments are worth looking at.
For instance, the TIMSS test~\cite{vonDavier24} includes tasks that require students to complete line charts by clicking. 

\begin{table*}
  \caption{\textbf{Example questions.}
  Statements that could be used to assess visualization interaction abilities, adapted from the work discussed in \autoref{sec:established}.}
  \label{tab:items}
  \begin{tabular}{p{61mm}p{21mm}p{25mm}p{20mm}p{33mm}}
    \toprule
    Adapted Statement & Answers & Related Concept & Population & Source\\
    \midrule
    ``I can create charts and graphs using a computer'' & Likert scale & Digital self-efficacy & School students & TIMSS 2023~\cite{vonDavier24} \\
    ``How would you describe your previous experiences with tablets?'' & Multiple-choice & Device literacy & General public & Study by Wyche et al.~\cite{wyche16} \\
    ``Why do you think the designer used the particular elements on this visualization?'' & Open-ended & Visual literacy & Children & Callow's framework~\cite{callow08} \\
    \bottomrule
  \end{tabular}
\end{table*}

\section{How To Assess Interactive Visualization Literacy?}

Based on the existing assessments and literacies discussed above, we thus see multiple approaches to assessing people's abilities to interact with visualizations.
A fundamental feature of our IVL model~\cite{leon26IVL} is that interactive visualization literacy is composed of multiple literacies. Therefore, we discuss criteria and techniques that could be applied to assess IVL, some of the multiliteracies, or the dimensions of any literacy.

\subsection{Criteria for Tests}

The nature of interactivity and how to (best) measure it introduces challenges to an interaction assessment; most notably, that it would need to assess a person's relation with a visualization \textit{over time}. Hence, we consider a few criteria that can be useful to discuss potential directions for assessing (see below). 

\begin{itemize}
\item[\textbf{C1}] \textbf{Assess \textit{high-level} thinking:} An ideal test would assess higher-level activities, e.g., on the strategic and tactical levels of the IVL model \cite{leon26IVL}. Likewise, standard literacy is not measured by writing individual letters (alone). 

\item[\textbf{C2}] \textbf{Technically simple:} An assessment test should also be technically simple so it can be run without major equipment and installations; ideally, any person would just be able to run the test on their device.

\item[\textbf{C3}] \textbf{Self-administered:} An assessment should also be able to be run and managed by the tested person with little or no assistance from an instructor.

\item[\textbf{C4}] \textbf{Scalable:} In the long term, we should be able to scale a test to many participants with minimal effort from an instructor, similar to existing tests for static visualizations.
\end{itemize}

These criteria might not be complete at all, but they point to two things: \textit{a)} that it might be hard to design a perfect interaction test, and that \textit{b)} there might be rather complementary methods balancing these criteria for specific scenarios.

\subsection{Assessment Techniques}

We summarize below what each approach to assessing visualization interaction abilities may entail, highlighting that none of these is probably able to satisfy all of the criteria above.

\begin{description}

    \item[Interviews and observations.]
    Asking open-ended questions with different sets of tasks adapted to the community, such as the proposals of Shreiner~\cite{shreiner19} and Callow~\cite{callow08}, could help generate insights into how and why people use interactive visualizations.
    Conducting interviews would also allow identifying barriers, characterizing the problems, and reflecting on how different literacy levels may look.
    The outcome would be qualitative insights that can help develop theoretical work further or guide the design of a scalable assessment.
    As an aside, this approach is similar to what we based our initial scrutiny of the IVL model on in our recent CHI 2026 paper~\cite{leon26IVL}.
    Beyond providing empirical validation of our model, we felt that an initial qualitative exploration would be the best way to understand how to assess the multiple literacies involved in interactive sensemaking.
    
    \item[Scale development based on Item Response Theory.]
    A different approach could draw from the visualization literature and expert surveys (similar to the methodology of Cabouat et al.~\cite{cabouat25}), to create and refine a list of test items (i.e., statement or question) that characterize interaction abilities.
    For example, based on the concepts described in \autoref{sec:interaction}, one could systematically develop test items in three categories (i.e., output capabilities, input capabilities, and framing).
    The outcome would be a test draft involving response items that take the use of standard interaction techniques into account, which would need to be validated through a user study.
    
    \item[Combining or extending existing assessments.]
    We could create a test for interactive visualization abilities by extending, combining, or adapting existing tests of related abilities (e.g., visualization literacy, digital literacy). Using established tests as building blocks (e.g., on problem-solving skills \cite{oecdadult25}, Mini-VLAT \cite{pandey23}) would serve as a ground truth to take fundamental abilities into account and determine whether and what additional response items could account for the aspects of visualization interaction that those tests do not cover.
    Exploratory or confirmatory analyses would help determine the dimensions or literacies that enable someone to interact.
    
    \item[Self-assessment.]
    It could be possible to combine the previous two approaches by asking people to self-assess their own perceived interaction ability.
    The outcome would be an assessment test with response items that would focus on asking people to assess their own skills; for example, a Likert scale to agree with the statement \textit{``I can use interactive charts to their full extent when I encounter them while reading the news.''} 
    Existing assessments on computer-related abilities could serve as a base for combining tasks or response items addressing related aspects (e.g., the TIMSS digital self-efficacy test).
    While people's judgement of their own abilities may differ from their actual abilities, such an assessment could cover scenarios that are challenging to recreate with fixed test tasks (e.g., how people transform a high-level goal into a series of tasks).

    \item[Biometric markers.] 
    Low-level biometric and physiological markers offer continuous, implicit measures that can be mapped to specific IVL literacies and gulfs.
    Eye tracking is the most mature modality for visualization research: scanpath analysis and fixation patterns can index visualization literacy during evaluation~\cite{DBLP:journals/tiis/SteichenCC14, DBLP:conf/icmi/BarralLGIC20}, while pupil dilation serves as a real-time proxy for cognitive load.
    Conati et al.\ showed that gaze data can predict cognitive abilities during visualization interaction with sufficient accuracy for user-adaptive systems~\cite{DBLP:journals/tiis/ConatiLRT20}.
    Galvanic skin response (GSR/EDA) captures arousal and stress responses that correlate with task difficulty~\cite{DBLP:conf/chi/ShiRTCC07, DBLP:conf/ozchi/Nourbakhsh0CC12}, potentially flagging breakdowns at specific gulfs---for example, EDA spikes preceding interaction abandonment could signal execution barriers.
    Mouse dynamics and interaction kinematics (hover hesitation, cursor velocity, click error rates) provide the cheapest continuous measures, indexing device and interaction literacy at the operational and tactical levels~\cite{DBLP:journals/tiis/ConatiLRT20}.
    A key opportunity lies in \textit{multimodal fusion} of these heterogeneous signals: combining eye tracking, physiological markers, and interaction logs with the multi-level, multi-gulf structure of the IVL model could enable attribution of specific biometric patterns to literacy deficits, moving beyond aggregate cognitive load measurement toward targeted diagnostics of visualization ability.

    \item[LLM coding of qualitative data.]
    Based on our initial empirical validation of the IVL model, we find that many of the literacies surveyed cannot easily be captured as quantitative measures.
    For example, assessing the reasoning involved in higher-level analytical literacies may require analyzing transcripts from a think-aloud protocol or free-text responses to analysis questions about a dataset.
    If we are to create any form of automated assessments at scale, we will thus have to turn to automatic coding and grading of such qualitative data.
    Recent years have seen several examples of using strictly controlled LLMs with specific parameter settings and prompts to achieve this; for example, Wang et al.~\cite{DBLP:journals/pacmhci/WangEBM25} demonstrated that setting the LLM temperature to zero yields reproducible thematic coding of qualitative interview data, while Hashemi et al.~\cite{DBLP:conf/acl/HashemiERDK24} showed that LLM-generated rubric assessments across multiple dimensions can be calibrated against human judges to achieve reliable evaluation.
    Multi-agent approaches such as AutoSCORE~\cite{DBLP:journals/corr/abs-2509-21910} decompose assessment into structured component recognition followed by item-by-item scoring, improving consistency on complex multidimensional rubrics---a structure that maps naturally onto multiliteracies \cite{newlondon1996multiliteracies}.
    A new assessment could leverage these techniques by designing per-literacy rubrics and feeding different data streams (e.g., think-aloud transcripts for goal formation literacies, interaction logs for execution literacies) to specialized prompts for automated scoring.

    \item[Interaction logging] Interaction logs would track a person's interaction with a system over a possibly long period of time: buttons clicked, time spent panning or zooming, time spent on specific views, etc. This would satisfy the criteria C2---C4, as any tracking would be implemented as part of the system itself without the intervention of a person. Logging could reveal if specific interaction features were used (or discovered) as well as how often they were used~\cite{molina22}. In an aggregated form, specific interaction patterns could point to tasks (tactical level) or even analysis patterns and goals (strategic level). However, logging will always suffer from a certain distance to the user, i.e., a lack of subjective and intentional information. The form of aggregating data may also imply the amount of high-level thinking trackable. For example, if certain analysis strategies can be formalized in terms of interaction, logging could help to explain them. 
\end{description}

Any of the techniques described requires choosing a target audience and context.
Large-scale assessments are often designed for school students and teachers (e.g., TIMSS~\cite{vonDavier24}), while visualization tests have been mostly tested with young adults (e.g., VLAT~\cite{lee17}).
The characteristics of the ability can also help favor or discard techniques; interviews and observations may be more suitable for high-level analysis abilities, while logging may fit best with interaction literacy~\cite{bach18}.
Hence, it is clear that no single assessment test may capture all literacies at once, that in fact one assessment test per literacy may be necessary, and that a final IVL rating would be a composite representing many such literacies.
While there is no right universal approach, each technique would lead to insights at different levels of scale that may help us understand better how people interact with data visualizations.

\section{Conclusion}
Assessing interaction abilities for data visualizations will likely require combining insights from multiple theoretical and measurement approaches. 
Whether through qualitative studies, standard response item development, or AI-assisted data processing, each approach has different advantages and challenges, illustrating one of multiple perspectives on how we can improve our understanding of interaction abilities. With this paper, we aim to contribute to the conversation on how multiple literacy concepts are interconnected and how we can benefit from the advancement of different fields to create robust assessment tools.


\bibliographystyle{ACM-Reference-Format}
\bibliography{assessing-interaction}

@String{jourCGF           = {Computer Graphics Forum}}

@String{jourIVS           = {Information Visualization}}

@String{jourPACM-HCI      = {Proceedings of the {ACM} on Human-Computer Interaction}}

@String{procCHI-EA        = {Extended Abstracts of the ACM Conference on Human Factors in Computing Systems}}

@String{procInfoVis       = {Proceedings of the {IEEE} Conference on Information Visualization}}

@String{pubACM            = {{ACM}}}

@String{pubIEEECS         = {{IEEE Computer Society}}}

@String{pubIEEECS         = {{IEEE CS}}}

@String{pubACL            = {Association for Computational Linguistics}}

@String{addrACL           = {Stroudsburg, PA, USA}}

@String{addrACM           = {New York, NY, USA}}

@String{addrIEEECS        = {Los Alamitos, CA, USA}}

@String{jourTVCG          = {{{IEEE} Transactions on Visualization and Computer Graphics}}}

@String{jourCGF           = {{Computer Graphics Forum}}}

@String{procCHI           = {Proceedings of the {ACM} Conference on Human Factors in Computing Systems}}

@String{procUIST          = {Proceedings of the {ACM} Symposium on User Interface Software and Technology}}

@String{pubACL            = {ACL}}

@inproceedings{DBLP:conf/ozchi/Nourbakhsh0CC12,
  author       = {Nargess Nourbakhsh and
                  Yang Wang and
                  Fang Chen and
                  Rafael A. Calvo},
  title        = {Using galvanic skin response for cognitive load measurement in arithmetic
                  and reading tasks},
  booktitle    = {Proceedings of the Australian Computer-Human Interaction Conference},
  pages        = {420--423},
  publisher    = pubACM, 
  address      = addrACM,
  year         = {2012},
  doi          = {10.1145/2414536.2414602},
}

@inproceedings{DBLP:conf/chi/ShiRTCC07,
  author       = {Yu (David) Shi and
                  Natalie Ruiz and
                  Ronnie Taib and
                  Eric H. C. Choi and
                  Fang Chen},
  title        = {Galvanic skin response {(GSR)} as an index of cognitive load},
  booktitle    = procCHI-EA,
  pages        = {2651--2656},
  publisher    = pubACM, 
  address      = addrACM,
  year         = {2007},
  doi          = {10.1145/1240866.1241057},
}

@article{DBLP:journals/tiis/ConatiLRT20,
  author       = {Cristina Conati and
                  S{\'{e}}bastien Lall{\'{e}} and
                  Md. Abed Rahman and
                  Dereck Toker},
  title        = {Comparing and Combining Interaction Data and Eye-tracking Data for
                  the Real-time Prediction of User Cognitive Abilities in Visualization
                  Tasks},
  journal      = {{ACM} Transactions on Interactive Intelligent Systems},
  volume       = {10},
  number       = {2},
  pages        = {12:1--12:41},
  year         = {2020},
  doi          = {10.1145/3301400},
}

@inproceedings{DBLP:conf/icmi/BarralLGIC20,
  author       = {Oswald Barral and
                  S{\'{e}}bastien Lall{\'{e}} and
                  Grigorii Guz and
                  Alireza Iranpour and
                  Cristina Conati},
  title        = {Eye-Tracking to Predict User Cognitive Abilities and Performance for
                  User-Adaptive Narrative Visualizations},
  booktitle    = {Proceedings of the ACM Conference on Multimodal Interaction},
  pages        = {163--173},
  publisher    = pubACM, 
  address      = addrACM,
  year         = {2020},
  doi          = {10.1145/3382507.3418884},
}

@article{DBLP:journals/tiis/SteichenCC14,
  author       = {Ben Steichen and
                  Cristina Conati and
                  Giuseppe Carenini},
  title        = {Inferring Visualization Task Properties, User Performance, and User
                  Cognitive Abilities from Eye Gaze Data},
  journal      = {{ACM} Transactions on Interactive Intelligent Systems},
  volume       = {4},
  number       = {2},
  pages        = {11:1--11:29},
  year         = {2014},
  doi          = {10.1145/2633043},
}

@article{DBLP:journals/corr/abs-2509-21910,
  author       = {Yun Wang and
                  Zhaojun Ding and
                  Xuansheng Wu and
                  Siyue Sun and
                  Ninghao Liu and
                  Xiaoming Zhai},
  title        = {AutoSCORE: Enhancing Automated Scoring with Multi-Agent Large Language
                  Models via Structured Component Recognition},
  journal      = {CoRR},
  volume       = {abs/2509.21910},
  year         = {2025},
  url          = {https://doi.org/10.48550/arXiv.2509.21910},
  doi          = {10.48550/ARXIV.2509.21910},
  eprinttype    = {arXiv},
  eprint       = {2509.21910},
  numpages     = {9},
}

@inproceedings{DBLP:conf/acl/HashemiERDK24,
  author       = {Helia Hashemi and
                  Jason Eisner and
                  Corby Rosset and
                  Benjamin Van Durme and
                  Chris Kedzie},
  title        = {{LLM-Rubric}: {A} Multidimensional, Calibrated Approach to Automated
                  Evaluation of Natural Language Texts},
  booktitle    = {Proceedings of the Annual Meeting of the Association for Computational
                  Linguistics},
  pages        = {13806--13834},
  publisher    = pubACL, 
  addr         = addrACL,
  year         = {2024},
  doi          = {10.18653/V1/2024.ACL-LONG.745},
  address = {Bangkok, Thailand}
}

@article{DBLP:journals/pacmhci/WangEBM25,
  author       = {Qile Wang and
                  Moath Erqsous and
                  Kenneth E. Barner and
                  Matthew Louis Mauriello},
  title        = {{LATA:} {A} Pilot Study on LLM-Assisted Thematic Analysis of Online
                  Social Network Data Generation Experiences},
  journal      = jourPACM-HCI,
  volume       = {9},
  number       = {2},
  pages        = {1--28},
  year         = {2025},
  doi          = {10.1145/3711022},
}

@inproceedings{Bederson1994,
    author = "Benjamin B. Bederson and James D. Hollan",
    title = "Pad++: A Zooming Graphical Interface for Exploring Alternate Interface Physics",
    booktitle = procUIST,
    year = {1994},
    publisher = pubACM,
    address = addrACM,
    doi = {10.1145/192426.192435},
    pages = {17--26}
}

@article{DBLP:journals/ivs/KandelHPKHRWLBB11,
  author       = {Sean Kandel and
                  Jeffrey Heer and
                  Catherine Plaisant and
                  Jessie Kennedy and
                  Frank van Ham and
                  Nathalie Henry Riche and
                  Chris E. Weaver and
                  Bongshin Lee and
                  Dominique Brodbeck and
                  Paolo Buono},
  title        = {Research directions in data wrangling: Visualizations and transformations for usable and credible data},
  journal      = jourIVS,
  volume       = {10},
  number       = {4},
  pages        = {271--288},
  year         = {2011},
  doi          = {10.1177/1473871611415994},
}

@inproceedings{DBLP:conf/infovis/Chi00,
  author       = {Ed Huai{-}hsin Chi},
  title        = {A Taxonomy of Visualization Techniques Using the Data State Reference Model},
  booktitle    = procInfoVis,
  pages        = {69--75},
  publisher    = pubIEEECS,
  address      = addrIEEECS,
  year         = {2000},
  doi          = {10.1109/INFVIS.2000.885092},
}

@article{DBLP:journals/tvcg/DimaraZTF22,
  author       = {Evanthia Dimara and
                  Harry Zhang and
                  Melanie Tory and
                  Steven Franconeri},
  title        = {The Unmet Data Visualization Needs of Decision Makers Within Organizations},
  journal      = jourTVCG,
  volume       = {28},
  number       = {12},
  pages        = {4101--4112},
  year         = {2022},
  doi          = {10.1109/TVCG.2021.3074023},
}

@book{DBLP:books/lib/Tukey77,
  author       = {John W. Tukey},
  title        = {Exploratory Data Analysis},
  publisher    = {Addison-Wesley},
  address      = {Boston, MA, USA},
  year         = {1977},
}

@inproceedings{DBLP:conf/infovis/AmarES05,
  author       = {Robert A. Amar and
                  James Eagan and
                  John T. Stasko},
  title        = {Low-Level Components of Analytic Activity in Information Visualization},
  booktitle    = procInfoVis,
  pages        = {111--117},
  publisher    = pubIEEECS,
  address      = addrIEEECS,
  year         = {2005},
  doi          = {10.1109/INFVIS.2005.1532136},
}

@article{DBLP:journals/computer/Shneiderman83,
  author       = {Ben Shneiderman},
  title        = {Direct Manipulation: {A} Step Beyond Programming Languages},
  journal      = {Computer},
  volume       = {16},
  number       = {8},
  pages        = {57--69},
  year         = {1983},
  doi          = {10.1109/MC.1983.1654471},
}

@inproceedings{pirolli2005sensemaking,
  author    = {Peter Pirolli and Stuart Card},
  title     = {The Sensemaking Process and Leverage Points for Analyst Technology as Identified Through Cognitive Task Analysis},
  booktitle = {Proceedings of the International Conference on Intelligence Analysis},
  year      = {2005},
  pages     = {2--4},
  url       = {https://andymatuschak.org/files/papers/Pirolli,%20Card%20-%202005%20-%20The%20sensemaking%20process%20and%20leverage%20points%20for%20analyst%20technology%20as.pdf},
  publisher = {MITRE Corporation},
  address   = {McLean, VA, USA}
}

@article{newlondon1996multiliteracies,
  author    = {{The New London Group}},
  title     = {A Pedagogy of Multiliteracies: Designing Social Futures},
  journal   = {Harvard Educational Review},
  OPTjournal   = {Harv. Educ. Rev.},
  volume    = {66},
  number    = {1},
  pages     = {60--92},
  year      = {1996},
  doi       = {10.17763/haer.66.1.17370n67v22j160u},
}

@book{card99,
  title={Readings in Information Visualization: Using Vision to Think},
  author={Card, Stuart K. and Mackinlay, Jock and Shneiderman, Ben},
  year={1999},
  publisher={Morgan Kaufmann},
  address = {San Francisco, CA, USA},
  isbn={9781558605336},
  lccn={98053660},
  series={Interactive Technologies},
  url={https://books.google.dk/books?id=wdh2gqWfQmgC},
}

@inproceedings{beaudouinlafon00,
author = {Beaudouin-Lafon, Michel},
title = {Instrumental interaction: an interaction model for designing post-{WIMP} user interfaces},
year = {2000},
publisher = pubACM,
address = addrACM,
url = {https://doi.org/10.1145/332040.332473},
doi = {10.1145/332040.332473},
booktitle = procCHI,
pages = {446--453},
}

@ARTICLE{yi07,
  author={Yi, Ji Soo and Kang, Youn ah and Stasko, John and Jacko, J.A.},
  journal=jourTVCG,
  title={Toward a Deeper Understanding of the Role of Interaction in Information Visualization},
  year={2007},
  volume={13},
  number={6},
  pages={1224--1231},
  doi={10.1109/TVCG.2007.70515},
  url = {https://doi.org/10.1109/TVCG.2007.70515}
}

@article{callow08,
author = {Callow, Jon},
title = {Show Me: Principles for Assessing Students' Visual Literacy},
journal = {The Reading Teacher},
volume = {61},
number = {8},
pages = {616-626},
doi = {10.1598/RT.61.8.3},
year = {2008}
}

@ARTICLE{boy14,
  author={Boy, Jeremy and Rensink, Ronald A. and Bertini, Enrico and Fekete, Jean-Daniel},
  journal=jourTVCG, 
  title={A Principled Way of Assessing Visualization Literacy}, 
  year={2014},
  volume={20},
  number={12},
  pages={1963-1972},
  doi={10.1109/TVCG.2014.2346984}
}

@inproceedings{wyche16,
author = {Wyche, Susan and Steinfield, Charles and Cai, Tian and Simiyu, Nightingale and Othieno, Martha E.},
title = {Reflecting on Video: Exploring the Efficacy of Video for Teaching Device Literacy in Rural Kenya},
year = {2016},
publisher = pubACM,
address = addrACM,
doi = {10.1145/2909609.2909667},
booktitle = {Proceedings of the ACM Conference on Information and Communication Technologies and Development},
OPTbooktitle = {Proc.\ ACM ICTD},
pages = {8:1--8:10},
}

@ARTICLE{lee17,
  author={Lee, Sukwon and Kim, Sung-Hee and Kwon, Bum Chul},
  journal=jourTVCG, 
  title={{VLAT}: Development of a Visualization Literacy Assessment Test}, 
  year={2017},
  volume={23},
  number={1},
  pages={551--560},
  doi={10.1109/TVCG.2016.2598920}
}

@inproceedings{bach18,
  title={Ceci n'est pas la data: Towards a Notion of Interaction Literacy for Data Visualization.},
  author={Bach, Benjamin},
  booktitle={Proceedings of AVI Workshop on Visual Interfaces for Big Data Environments in Industrial Applications},
  pages={1--3},
  publisher = pubACM,
  address = addrACM,
  year={2018},
  url={https://ceur-ws.org/Vol-2108/invited1.pdf},
}

@inproceedings{ge23,
author = {Ge, Lily W. and Cui, Yuan and Kay, Matthew},
title = {{CALVI}: Critical Thinking Assessment for Literacy in Visualizations},
year = {2023},
  publisher = pubACM,
  address = addrACM,
  doi = {10.1145/3544548.3581406},
  booktitle = procCHI,
  pages = {815:1--815:18},
}

@article{pandey23,
  author = {Pandey, Saugat and Ottley, Alvitta},
  title = {Mini-{VLAT}: A Short and Effective Measure of Visualization Literacy},
  journal = jourCGF,
  volume = {42},
  number = {3},
  pages = {1--11},
  doi = {10.1111/cgf.14809},
  year = {2023}
}

@ARTICLE{cabouat25,
  author={Cabouat, Anne-Flore and He, Tingying and Isenberg, Petra and Isenberg, Tobias},
  journal=jourTVCG, 
  title={{PREVis}: Perceived Readability Evaluation for Visualizations}, 
  year={2025},
  volume={31},
  number={1},
  pages={1083--1093},
  doi={10.1109/TVCG.2024.3456318}
}

@article{carolus23,
title = {Digital interaction literacy model – Conceptualizing competencies for literate interactions with voice-based AI systems},
journal = {Computers and Education: Artificial Intelligence},
volume = {4},
pages = {100114},
year = {2023},
issn = {2666-920X},
doi = {https://doi.org/10.1016/j.caeai.2022.100114},
url = {https://www.sciencedirect.com/science/article/pii/S2666920X22000698},
author = {Astrid Carolus and Yannik Augustin and André Markus and Carolin Wienrich},
}

@techreport{vonDavier24,
  author    = {von Davier, Matthias and Kennedy, Ann and Reynolds, Katherine and Fishbein, Bethany and Khorramdel, Lale and Aldrich, Charlotte and Bookbinder, Allison and Bezirhan, Ummugul and Yin, Liqun},
  title     = {{TIMSS} 2023 International Results in Mathematics and Science},
  institution = {Boston College, TIMSS \& PIRLS International Study Center},
  year      = {2024},
  doi       = {10.6017/lse.tpisc.timss.rs6460},
  url       = {https://doi.org/10.6017/lse.tpisc.timss.rs6460}
}

@techreport{oecdadult25,
    author = {OECD},
    title = {Survey of Adult Skills 2023 Technical Report},
    institution = {OECD Skills Studies},
    year = {2025},
    publisher = {OECD Publishing},
    address = {Paris},
  doi       = {10.1787/80d9f692-en},
  url       = {https://doi.org/10.1787/80d9f692-en}
}

@article{cui23,
author = {Ying Cui and Fu Chen and Alina Lutsyk and Jacqueline P. Leighton and Maria Cutumisu},
title = {Data literacy assessments: a systematic literature review},
journal = {Assessment in Education: Principles, Policy \& Practice},
volume = {30},
number = {1},
pages = {76--96},
year = {2023},
publisher = {Routledge},
doi = {10.1080/0969594X.2023.2182737},
URL = {https://doi.org/10.1080/0969594X.2023.2182737},
}

@article{shreiner19,
author = {Tamara L. Shreiner},
title ={Students’ Use of Data Visualizations in Historical Reasoning: A Think-Aloud Investigation with Elementary, Middle, and High School Students},
journal = {The Journal of Social Studies Research},
volume = {43},
number = {4},
pages = {389-404},
year = {2019},
doi = {10.1016/j.jssr.2018.11.001},
URL = {https://doi.org/10.1016/j.jssr.2018.11.001},
}

@Article{lee19corr,
AUTHOR = {Lee, Sukwon and Kwon, Bum Chul and Yang, Jiming and Lee, Byung Cheol and Kim, Sung-Hee},
TITLE = {The Correlation between Users’ Cognitive Characteristics and Visualization Literacy},
JOURNAL = {Applied Sciences},
VOLUME = {9},
YEAR = {2019},
NUMBER = {3},
ARTICLE-NUMBER = {488},
URL = {https://www.mdpi.com/2076-3417/9/3/488},
DOI = {10.3390/app9030488},
numpages = {20},
articleno = {488},
}

@article{molina22,
author = {Molina León, Gabriela and Lischka, Michael and Luo, Wei and Breiter, Andreas},
title = {Mobile and Multimodal? {A} Comparative Evaluation of Interactive Workplaces for Visual Data Exploration},
journal = jourCGF,
volume = {41},
number = {3},
pages = {417--428},
url = {https://doi.org/10.1111/cgf.14551},
year = {2022}
}

@preprint{leon26IVL,
  title={A Multiliteracy Model for Interactive Visualization Literacy: Definitions, Literacies, and Steps for Future Research},
  author={Molina León, Gabriela and Bach, Benjamin and Valentim, Matheus and Elmqvist, Niklas},
  year={2026},
  archivePrefix = {arXiv},
  eprint = {2602.09631},
  DOI={10.1145/3772318.3793423},
}


\end{document}